\documentclass{svjour3}                     
\smartqed  
\usepackage{graphicx}
\usepackage[T1,hyphens]{url}
\usepackage{hyperref} 
\usepackage{amsmath}
\usepackage{subcaption}
\usepackage{xcolor}

\graphicspath{{figs/}} 

\journalname{SN Computer Science}
\begin{document}

\title{Examination of Community Sentiment Dynamics due to COVID-19 Pandemic: A Case Study from A State in Australia
\thanks{This paper has been published by SN Computer. Please cite this paper with: J Zhou, S Yang, C Xiao, F Chen, Examination of Community Sentiment Dynamics due to COVID-19 Pandemic: A Case Study from a State in Australia, SN Computer Science 2 (3), 1-11, 2021}}


\titlerunning{Examination of Community Sentiment Dynamics}        

\author{Jianlong Zhou$^{*}$ 
        \and
        Shuiqiao Yang 
        \and \\
        Chun Xiao
        \and
        Fang Chen
}


\institute{$^{*}$Corresponding author \\
J. Zhou \at
              Data Science Institute\\
              University of Technology Sydney, Australia\\
              \email{jianlong.zhou@uts.edu.au}           
           \and
           S. Yang \at
              Data Science Institute\\
              University of Technology Sydney, Australia \\ 
              \email{shuiqiao.yang@uts.edu.au} 
            \and
           C. Xiao \at
              Faculty of Transdisciplinary Innovation\\
              University of Technology Sydney, Australia\\
              \email{chun.xiao@uts.edu.au}
            \and
           F. Chen \at
              Data Science Institute\\
              University of Technology Sydney, Australia\\
              \email{fang.chen@uts.edu.au}
}

\date{Received: date / Accepted: date}

\maketitle

\begin{abstract}


The outbreak of the novel Coronavirus Disease 2019 (COVID-19) has caused unprecedented impacts to people's daily life around the world. Various measures and policies  such as lockdown and social-distancing are implemented by governments to combat the disease  during the pandemic period. These measures and policies as well as virus itself may cause different mental health issues to people such as depression, anxiety, sadness, etc.
In this paper, we exploit the massive text data posted by Twitter users to analyse the sentiment dynamics of people living in the state of New South Wales (NSW) in Australia during the pandemic period. 
Different from the existing work that mostly focuses the country-level and static sentiment analysis, 
we analyse  the sentiment dynamics at the fine-grained local government areas (LGAs).
Based on the analysis of around 94 million tweets that posted by around 183 thousand users located at different LGAs in NSW in five months, we found that people in NSW showed an overall positive sentimental polarity and the COVID-19 pandemic decreased the overall positive sentimental polarity during the pandemic period.
The fine-grained analysis of sentiment in LGAs found that despite the dominant positive sentiment most of days during the study period, some LGAs experienced significant sentiment changes from positive to negative.
This study also analysed the sentimental dynamics delivered by the hot topics in Twitter such as government policies (e.g. the Australia's JobKeeper program, lockdown, social-distancing) as well as the focused social events (e.g. the Ruby Princess Cruise). The results showed that the policies and events did affect people's overall sentiment, and they affected people's overall sentiment differently at  different stages.



\end{abstract}

\section{Introduction}
\label{sec:intro}



The World Health Organization (WHO) declared novel Coronavirus Disease 2019 (COVID-19) as a pandemic in March 2020 \cite{noauthor_coronavirus_2020}. It has now spread across the world and there have been more than 6.29 million confirmed cases and more than 380,000 people died because of the virus until 4 June 2020. Almost all countries in the world are battling against COVID-19 to prevent it from the spread as much as possible. 
The impact of the COVID-19 outbreak has been so huge that it is said to be the most serious epidemic in the past hundred years comparable to pandemics of the past like Spanish flu of 1918, or the Black Death in the mid-1300s. 
The outbreak has caused scares across the globe affecting millions of people either through infection or through increased mental health issues, such as disruption, stress, worry, fear, disgust, sadness, 
anxiety (a fear for one’s own health,and a fear of infecting others), and perceived stigmatisation \cite{montemurro_emotional_2020,bhat_sentiment_2020}.
These mental health issues can even occur in people not at high risk of getting infected, in the face of a virus with which the common people may be unfamiliar \cite{montemurro_emotional_2020}. 
These mental health issues can cause severe emotional, behavioral and physical health problems. Complications linked to mental health include:
unhappiness and decreased enjoyment of life, family conflicts, relationship difficulties, social isolation, problems with tobacco, alcohol and other drugs, self-harm and harm to others (such as suicide or homicide), weakened immune system, and heart disease as well as other medical conditions.

Furthermore, lockdown is a commonly completed measure to stop the spread of virus in the community. For example, countries like China, Italy, Spain, and Australia were fighting with the COVID-19 pandemic by taking strict measures of national-wide lockdown or by cordoning off the areas that were suspected of having risks of community spread throughout the period of the pandemic, expecting to  ``flatten the curve''. 
Therefore, it is critical to learn the status of the community mental health especially during the pandemic period so that corresponding measures can be taken in time.

On the other hand, social media such as Twitter represent a relatively real-time large-scale snapshot of messages reflecting people's thoughts and feelings. Every tweet is a signal of the social media users' state of mind and state of being at that moment \cite{gibbons_twitter-based_2019}. Aggregation of such digital traces may make it possible to monitor health behaviours at a large-scale \cite{jaidka_estimating_2020}.
While during the lockdown of COVID-19, people have taken social media to express their feelings and find a way to calm themselves down. Therefore, social media have created possibilities of analysing people's sentiment and its dynamics during the pandemic period. The most recent work have used Twitter to analyse public sentiment during the pandemic \cite{barkur_sentiment_2020,bhat_sentiment_2020}. These work largely focus on the public sentiment in country-level or even multiple countries for a specific period. However, a granular analysis of sentiment of a single city or even a suburb is more operable for related parties such as governmental departments to take corresponding actions. Furthermore, the sentiment dynamics over time instead of sentiment of a specific period will help related parties understand the effectiveness of any measures implemented.

This paper aims to examine community sentiment dynamics due to COVID-19 pandemic in Australia. Twitter data are collected and analysed to extract sentiment scores which may be affected by COVID-19 and related events during the COVID-19 period. Instead of the sentiment examination of the whole country, this paper investigates a fine-grained analysis of sentiment in local government areas of a state in Australia, which can help government to take corresponding measures more objectively if necessary. Furthermore, through analysing topic key words labelled by hashtag in Twitter, this paper examines community dynamics because of different events, government policies and programs or others with visual analytics.



\section{Related Work}
\label{sec:relatedwork}


\subsection{Social media sentiment analysis}
Sentiment analysis is a research topic that analyzes people's sentiments, opinions, mental states, and emotions from data resources like texts, images and videos that generated by human beings \cite{Liu2012}. 
With the rapidly development of social media platforms such as Twitter and Facebook \cite{Yang2020}, it becomes possible to collect large-scale of text data from millions of users for the social medial sentiment analysis \cite{Anuprathibha2016}.  
Most of the methods take the concept of supervised machine learning for social media sentimental analysis \cite{Go2009,Mittal2012,Jianqiang2018,Hassan2013}. 
In those methods, the supervised machine learning models classify a tweet into predefined sentimental categories such as positive and negative. Specifically, the supervised machine learning models firstly need to be trained with data that contain the labels for different sentiments. Then, the trained models can be used for sentiment prediction.  
For instance, Go et al. \cite{Go2009} adopted the distant supervision strategy to create a labeled tweet dataset for training supervised machine learning models such as Naive Bayes, Maximum Entropy, and support vector machine. The emoticons such as :), :( are served as noise labels to indicate the sentiment of the tweet publisher as positive or negative. With the superior advantages of automatically feature selection from raw data, deep learning techniques have also been adopted for sentiment analysis. Jiang et al. \cite{Jianqiang2018} adopted the the latest techniques of deep neural network as the sentimental classifier for more accurate sentiment classification. 
Except the machine learning based methods for sentiment analysis, 
there some lexicon-based methods that exploiting the word polarity to compute the sentimental score for a given text \cite{hutto_VADER_2014,thelwall2010sentiment,ortega2013ssa}.
For example, Ortega et al. \cite{ortega2013ssa} proposed to exploit multiple steps of different techniques for Twitter sentimental analysis. Firstly, the pre-processing for tweets was performed to clean the raw texts. Then, an existing polarity detection method were adopted to compute the sentimental score of each word. Finally, they adopted a rule-based classification to classify the sentiments for tweets. The adopted polarity detection and rule-based classification methods were based on existing knowledge bases: WordNet and SentiWordNet. 

\subsection{Sentiment analysis for COVID-19 based on Twitter}
More recently, with the pandemic of COVID-19 spreading in the world, there are some work that exploiting sentiment analysis as a tool to investigate the people's reactions. For instance,  
Barkur et al. \cite{barkur_sentiment_2020} analysed public sentiment of Indians after the lockdown announcements were made. The social media platform Twitter was used for the sentiment analysis.
The analysis showed that Indians have taken the fight against COVID-19 positively and majority are in agreement with the government for announcing the lockdown to flatten the curve. However, this work focused on the overall public sentiment of a specific period but did not show the dynamics of sentiment during the study period.
Bhat et al. \cite{bhat_sentiment_2020} analysed sentiments expressed globally through Twitter and found that the perception of Twitter users was mostly positive or neutral during the COVID-19 pandemic period. It indicated that even though Twitter users were quarantined or staying at home, yet they were hopeful and experiencing a different unique socialization opportunity with family. 
Dubey \cite{dubey_twitter_2020} analysed tweets from twelve countries during a selected period of the pandemic, aiming to understand how the citizens of different countries are dealing with the situation. 
The results showed that while majority of people were taking a positive and hopeful approach, there were instances of fear, sadness and disgust exhibited worldwide. Furthermore, four countries, France, Switzerland, the Netherland and USA have shown signs of distrust and anger on a bigger scale as compared to the remaining eight countries. 

However, these previous work on COVID-19 sentiment analysis largely focuses on the public sentiment in country-level or globally for a specific period.  However, little work is found to analyse dynamics of community sentiment over time for different regions especially when government policies are applied or significant events are occurred during the pandemic period.

\section{Data and Methods}

\subsection{Study Location}

For this study, we focus on the state New South Wales (NSW) in Australia. NSW has a large population of around 8.1 million based on the census in September 2019 from Australian Bureau of Statistics\footnote{\url{https://www.abs.gov.au/}}. The state's capital city Sydney is Australia's most populated city with a population of over 5.3 million people. 
The Local Government Areas (LGAs) of New South Wales are the third tier of government in the Australian state (the three tiers are federal, state, and local government). There are 128 LGAs in NSW. In this study, the Twitter data were collected for each LGA separately so that the sentiment dynamics can be analysed and compared among LGAs.

\subsection{Measuring Community Sentiment}

Sentiment analysis, also known as opinion mining or emotion AI, is a sub-field of Natural Language Processing (NLP) that tries to identify, extract, quantify, and study affective states and subjective information within a given text.
We utilize the VADER (Valence Aware Dictionary for sEntiment Reasoning) \cite{hutto_VADER_2014} to analyse sentiments implied in tweets. VADER is a lexicon and rule-based sentiment analysis tool that is specifically attuned to sentiments expressed in social media. 

In this study, VADER is used to compute the sentiment value of each tweet. The sentiment of all tweets in an LGA is then aggregated for each LGA. 
In this study, sentiment of each LGA is aggregated with the Eq. \ref{equ:sentiment_agg1} - \ref{equ:sentiment_agg3} to get the dominant mean sentiment of LGA, which is also called community sentiment:

\begin{align}
\label{equ:sentiment_agg1}
    S_p = \frac{\sum_{i}^{n_p}s_{p}^{i}}{N}\\
    S_g = \frac{\sum_{i}^{n_g}s_{g}^{i}}{N}\\
    S_f = \left\{\begin{matrix}
    \label{equ:sentiment_agg3}
        S_p, & if \left | S_p \right |>=  \left | S_g \right |\\ 
        S_g, & if \left | S_p \right |< \left | S_g \right |
    \end{matrix}\right.
\end{align}

\noindent where $s_{p}^{i}$ is the sentiment value of tweet $i$ which has positive sentiment, $n_p$ is the number of tweets which have positive sentiment, 
where $s_{g}^{i}$ is the sentiment value of tweet $i$ which has negative sentiment, $n_g$ is the number of tweets which have negative sentiment, $N$ is the number of all tweets in an LGA, and $S_f$ is the final sentiment of an LGA.

Table \ref{tab:positive_tweets} and Table \ref{tab:negative_tweets} show examples of tweets with positive and negative sentiment values computed with VADER.



\begin{table}[!htb]
   \caption{Examples of tweets with positive sentiment.}
    \centering
    \begin{tabular}{l|l}
    \hline
    \textbf{Tweets} & \textbf{Sentiment}\\
    \hline
    Thank you all for hanging out in stream tonight. It was a lot of fun! ...  & \\ 
    for the amazing support you are all amazing. & 0.9628\\
    \hline
this interesting piece by the team at Macrobusiness suggests that it &\\
mainly benefits the wealthy, and the working classes ... & 0.9606\\
\hline
Harold is my best friend, and he's got a personality unlike any other cat ... &\\
I hope the other cats of the litter have gone on to brighten the lives of ... &0.967\\
\hline
Congratulations! He’s adorable. I hope you’re all safe and well and &\\
getting some sleep & 0.9652\\
\hline
Good evening my dear Bianca, another beautiful day here. Hope you are &\\
enjoying the sunshine too.  Stay safe ...& 0.9869\\
\hline
you are literally the best person! who would ever thing of making a zoom &\\
release party with fans? oh yeah, YOU! you are the most kind and caring ...& 0.9768\\
\hline
So great to see the quality care happening  at MHOC Vacation Care. & \\
Happy kids, fun activities, safe environment & 0.9896\\
         \hline
    \end{tabular}
    \label{tab:positive_tweets}
\end{table}

\begin{table}[!htb]
   \caption{Examples of tweets with negative sentiment.}
    \centering
    \begin{tabular}{l|l}
    \hline
    \textbf{Tweets} & \textbf{Sentiment}\\
    \hline
   So between this and CO repeating it's just a flu and the rest is a dem hoax, &\\ 
   WHY THE BLOODY HELL BLAME CHINA?? HAD THEY BLOODY ... & -0.9741\\
   \hline
So sad it's painful to see \#CoronaVirusinKenya has become a weapon to &\\
hurt our people. God be our shield & -0.9604\\
\hline
The most overweight liar? Tells the biggest lies? The most dangerous lies? &\\ 
The most lies per word spoken? Or all four? & -0.9623\\
\hline
SARS-2+A NEW AGE; SARS-2 (COVID19) is an opportunity to stop &\\
all the lies from publicly paid servants and fraudulent statement... & -0.9661\\
\hline
Guns at Protests, Looks More like a Threat of Violence... These are the &\\ 
Crazy Bastards Dangerous Bastards... \#MichiganProtest \#Michigan & -0.9655\\
\hline
Yea, criminals usually don't like anyone telling them that they can't be &\\
criminals. Fired Up? Well we’re fucking fired up too. So fuck those ... & -0.9762\\
\hline
Find them, charge them for assault and braking social distancing laws. &\\
There is no excuse for this! Shameful, disgusting, ... behaviour! & -0.9702\\
         \hline
    \end{tabular}
    \label{tab:negative_tweets}
\end{table}

\subsection{Analytical Approach}

In this study, various analytical approaches are applied to examine community sentiment dynamics before the outbreak of COVID-19, during the COVID-19 period, and post period of the COVID-19. The community sentiment may be affected by various significant events or policies implemented by the government during the COVID-19 period such as the state lockdown or restrictions to movement.  

The community sentiment of overall NSW state is firstly aggregated and analysed. We then split the community sentiment analysis for each LGA of NSW to see the sentiment difference among LGAs. The daily and weekly dynamics of these community sentiment are examined to find how various factors affect community sentiment. 

\subsection{Datasets}


In order to analyse the dynamics of sentiment during the COVID-19 pandemic period in a fine-grained level, we collected tweets from Twitter users that live in the different LGAs of NSW in Australia. 
The time span of the collected tweets is from 1 January 2020 to 22 May 200 which covers dates that the first confirmed case of coronavirus was reported in NSW (22 January 2020) and the first time that the NSW premier announced the relaxing for the lockdown policy (10 Mary 2020).
Table \ref{tab:lga_user_num} shows the statistics of the collected tweet dataset. We totally collected 94,707,264 tweets with averagely 739,901 tweets for each LGA during the study period.
Datasets of COVID-19 tests and confirmed cases in NSW were collected from DATA.NSW \footnote{\url{https://data.nsw.gov.au/}}.

\begin{table}[!htb]
   \caption{Statistics of the collected Twitter dataset.}
    \centering
    \begin{tabular}{l|l}
    \hline
    \textbf{Description} & \textbf{Numbers}\\
    \hline
    Total Twitter users &  183,104\\
         Average Twitter user per LGA & 1,430.5  \\ 
         Average tweets per LGA& 739,900.5\\
         Total tweets & 94,707,264\\
         \hline
    \end{tabular}
    \label{tab:lga_user_num}
\end{table}

\section{Results}

\begin{figure*}[t!]
    \centering
    \begin{subfigure}[t]{0.99\textwidth}
        \centering
        \includegraphics[width=\linewidth]{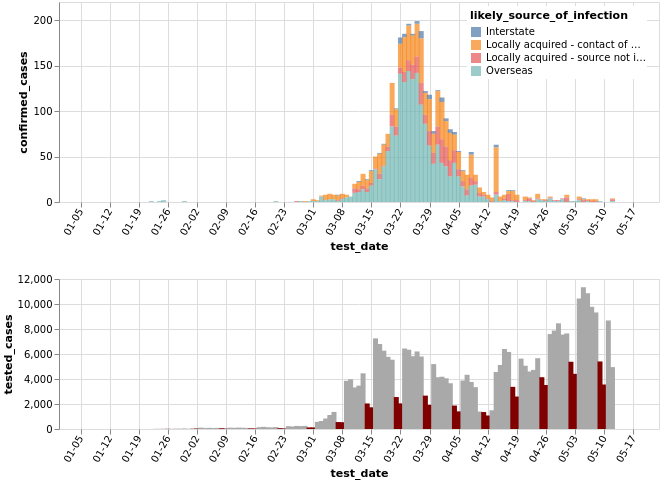}
        \caption{The tests and confirmed cases of COVID-19 in NSW.}
        \label{fig:nsw_test_confirmed_cases}
    \end{subfigure}\\%
    \begin{subfigure}[t]{0.98\textwidth}
        \centering
        \includegraphics[width=\linewidth]{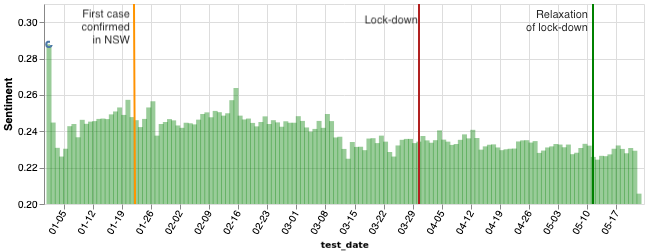}
        \caption{The overall community sentiment dynamics in NSW during the study period.}
        \label{fig:nsw_overall_sentiment}
    \end{subfigure}
    \caption{The COVID-19 spread and community sentiment in NSW.}
    \label{fig:nsw_cases_sentiment}
\end{figure*}

Fig. \ref{fig:nsw_test_confirmed_cases} shows the overview of the number of tests and confirmed cases of COVID-19 in NSW over the COVID-19 period. It was found that there usually had test peaks at the beginning of each week and had less numbers at the weekend, which well aligns with the people's living habits in Australia. It shows that the outbreak peak of COVID-19 in NSW was on 26 March 2020 and tests were significantly increased after 13 April 2020. It also shows that most of confirmed cases were originally related to overseas.


\subsection{Sentiment Dynamics in NSW}

Fig.~\ref{fig:nsw_overall_sentiment} presents the overall community sentiment dynamics in NSW during the study period. It shows that the public had a highly positive sentiment on the new year day. There was a significant decrease until 5 January. This was maybe because of the continued bushfire in NSW at that period. After that, the sentiment of the public was increased and then kept relatively stable until 8 March. From 8 March, the overall sentiment was decreased significantly and kept in low level continuously. This is well aligned with the overall trend of COVID-19 spreading in NSW as shown in Fig.~\ref{fig:nsw_test_confirmed_cases}, which shows that the number of confirmed cases of COVID-19 was increased significantly from 8 March. Overall, the community showed a dominant positive sentiment during the study period despite the COVID-19 spread. This observation is also aligned with findings from other researchers who got the similar conclusions from one country or several countries \cite{barkur_sentiment_2020,bhat_sentiment_2020}. Furthermore, it shows that the COVID-19 spread did affect community sentiment and decreased the community sentiment during the COVID-19 pandemic, which can be clearly seen from Fig.~\ref{fig:nsw_test_confirmed_cases}.     


\subsection{Sentiment Dynamics in LGAs}

\begin{figure}[!htb]
\centering
  \includegraphics[width=0.8\linewidth]{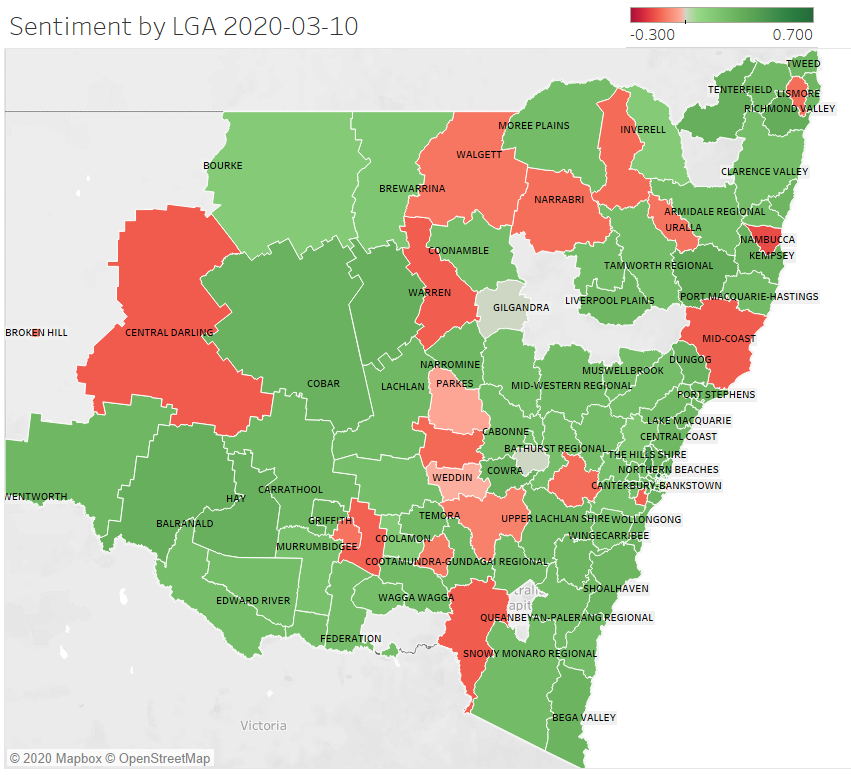}\\
  \includegraphics[width=0.8\linewidth]{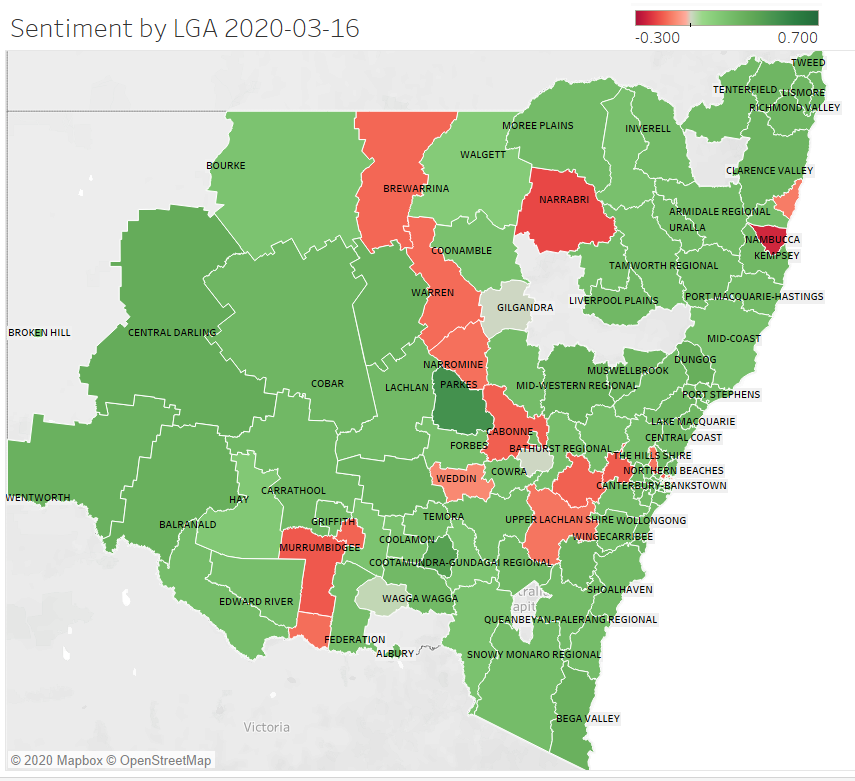}
  \caption{The community sentiment map in LGAs in NSW on 10 March 2020 (top) and 16 March 2020 (bottom).}
\label{fig:lga_sentiment1_all}
\end{figure}

\begin{figure}[!htb]
  \includegraphics[width=0.45\linewidth]{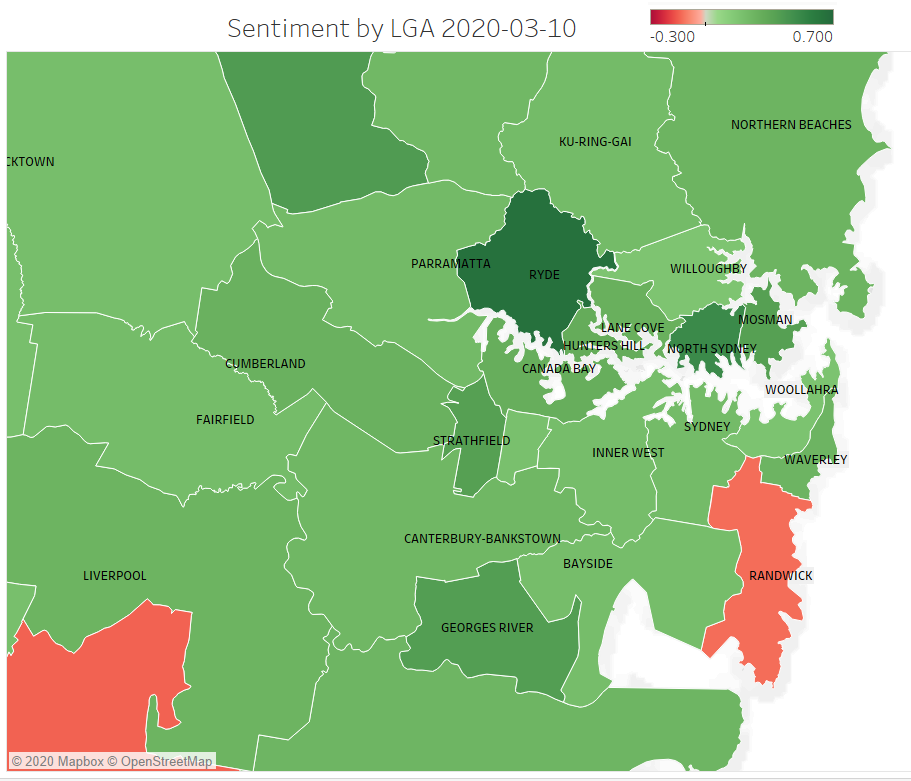}
  \includegraphics[width=0.5\linewidth]{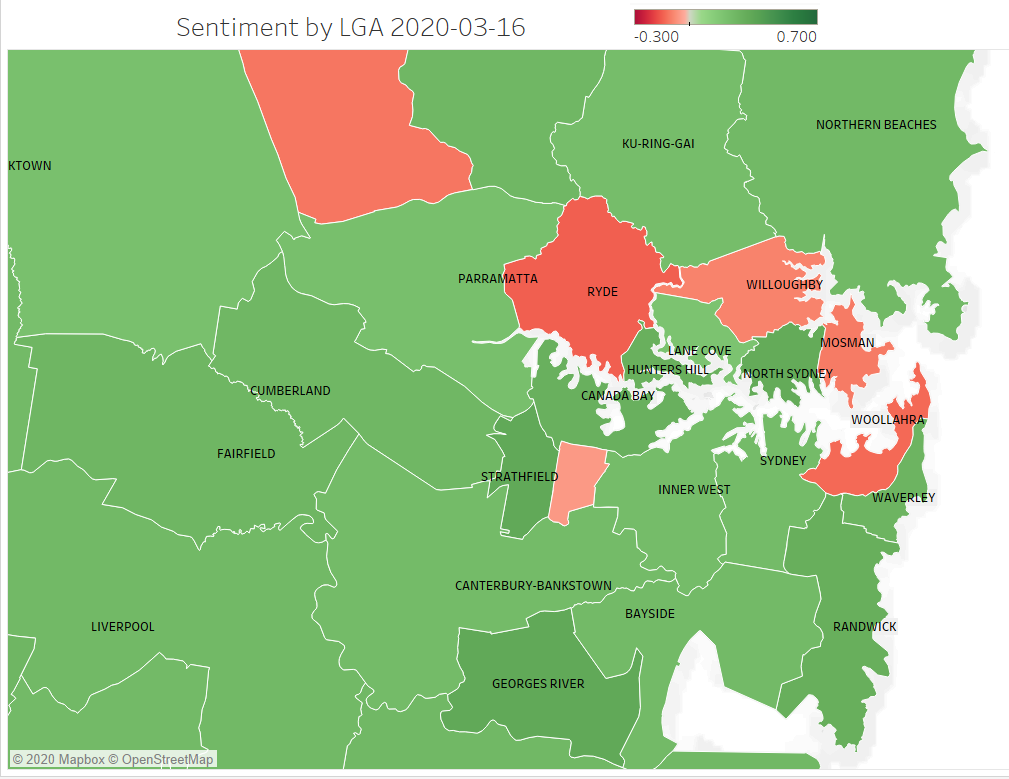}
  \caption{The community sentiment dynamics in selected LGAs around Sydney City areas on 10 March 2020 (left) and 16 March 2020 (right).}
\label{fig:lga_sentiment_selected}
\end{figure}

This subsection further analyses community sentiment dynamics in LGAs in NSW. Fig.~\ref{fig:lga_sentiment1_all} shows an example of community sentiment dynamics in LGAs in NSW on different two days: 10 March 2020 and 16 March 2020. It shows that the community sentiment were different across different LGAs on each day. Furthermore, the community sentiment of each LGA was changed on different days. When we zoom in the sentiment map to LGAs around Sydney City areas as shown in Fig.~\ref{fig:lga_sentiment_selected}, we can see more details of sentiment changes of LGAs around Sydney City areas on these two days. For example, Ryde is an LGA close to Sydney City. In an aged care centre in this LGA, a nurse and an 82-year-old elderly resident were tested positive for coronavirus at the beginning of March\footnote{\url{https://www.smh.com.au/national/woman-catches-coronavirus-in-australia-40-sydney-hospital-staff-quarantined-20200304-p546lf.html}}. After that, a number of elderly residents in this aged care centre were tested positive for COVID-19 or even died. At the same time, a childcare centre and a hospital in this LGA have been reported positive COVID-19 cases in March. Many staff from the childcare centre and the hospital were asked to test virus and conduct home isolation for 14 days. All these maybe resulted in public sentiment changes significantly as shown in Fig.~\ref{fig:lga_sentiment_selected}. For example, the sentiment in Ryde LGA was changed from high positive to high negative on 10 March 2020 and 16 March 2020. 

\begin{figure}[!htb]
  \includegraphics[width=0.99\linewidth]{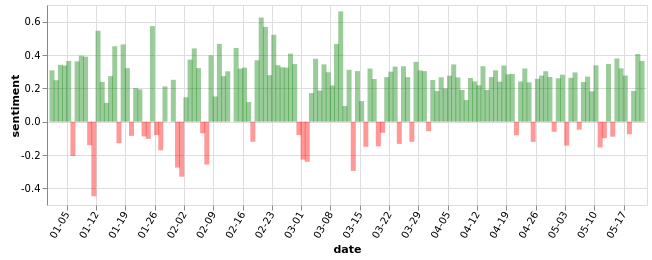}
  \caption{The community sentiment dynamics in Ryde LGA.}
\label{fig:lga_sentiment_ryde}  
\end{figure}

Fig.~\ref{fig:lga_sentiment_ryde} presents the community sentiment dynamics in Ryde LGA over the study period. It shows that the community experienced frequent negative sentiment in March, and recovered to positive dominantly in April. This is maybe because the COVID-19 spread situation in March as mentioned previously in this LGA and the spread was controlled in April, which helped people boost positive sentiment.

\subsection{Dynamics of Twitter Topics}

\begin{figure}[!htb]
  \includegraphics[width=0.99\linewidth]{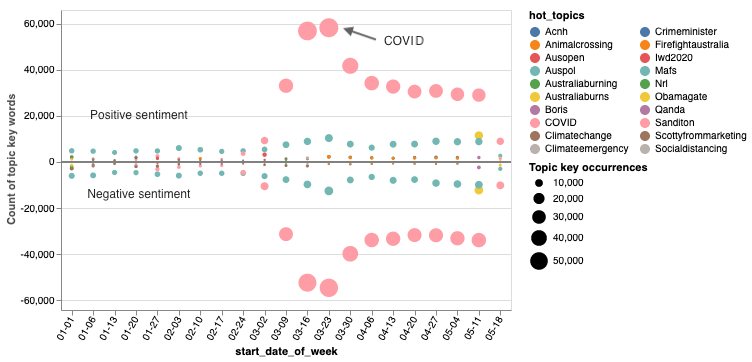}
  \caption{The dynamics of top 20 Twitter topics during the research period.}
\label{fig:HT_top20}
\end{figure}

We analysed hot topics in tweets labelled with hashtag. The sentiment of each tweet is got using the VADER tool as described previously. If the sentiment of a tweet is positive, the topic key words labelled with hashtag in this tweet are then labelled as positive sentiment. The topic key words with the negative sentiment are got with the similar approach. Different topic key words are then analysed statistically to understand the sentiment dynamics. Fig.~\ref{fig:HT_top20} shows the dynamics of top 20 Twitter topics during the study period in NSW. Y-axis represents the count of different topic key words occurred in tweets. Positive numbers represent count of topic key words with positive sentiment, and negative numbers represent count of topic key words with negative sentiment. This is same for other figures from Fig.~\ref{fig:HT_lockdown} to Fig.~\ref{fig:HT_rubyprincess}. 

\subsubsection{Topic of COVID-19}

Fig.~\ref{fig:HT_top20} shows that COVID is the dominant topic in the study period from the week of 2 March. When we drill down into details of this figure, we can see that the occurrence of the topic of COVID appeared from the week of 20 January, and began to increase significantly from the week of 2 March. It reached the peak in the week of 23 March. After that, the occurrence of the topic of COVID decreased gradually and kept stable from the week of 20 April. This trend is well aligned with the trend of COVID-19 situation in NSW as shown in Fig.~\ref{fig:nsw_test_confirmed_cases}. Furthermore, Fig.~\ref{fig:HT_top20} shows that the dominant sentiment around the topic of COVID was positive. This conclusion is aligned with findings in previous research \cite{barkur_sentiment_2020,bhat_sentiment_2020}. This is maybe because that people had a positive belief to the combat against COVID-19 by the government and the society.

\subsubsection{Topic of lockdown}
\begin{figure}[!htb]
  \includegraphics[width=0.99\linewidth]{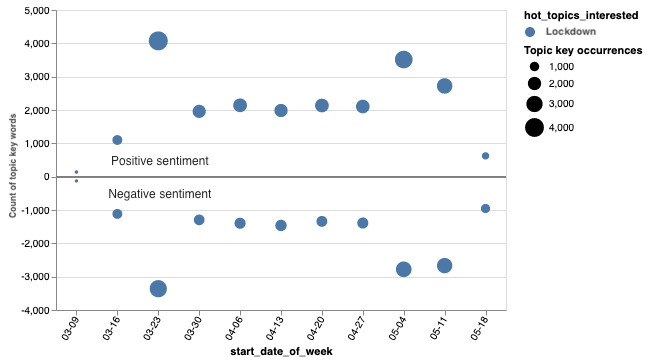}
  \caption{The dynamics of Twitter topic of ``lockdown''.}
\label{fig:HT_lockdown}
\end{figure}

Fig.~\ref{fig:HT_lockdown} shows the dynamics of Twitter topic of lockdown. From this figure, we can see that the topic of lockdown was started from the week of 9 March and reached to the peak point at the week of 23 March. On that week, NSW government officially announced the state lockdown on 30 March and the restrictions were begun from 31 March\footnote{\url{https://gazette.legislation.nsw.gov.au/so/download.w3p?id=Gazette_2020_2020-65.pdf}
}. After that, this topic was kept stable until the week of 27 April. From the week of 4 May, there was a small peak of the topic occurrences. Maybe this was because that the NSW government announced the ease of restrictions on 10 May\footnote{\url{https://www.nsw.gov.au/media-releases/nsw-to-ease-restrictions-week-0}}. Overall, the community kept a positive sentiment dominantly to lockdown despite negative sentiment existing. This was maybe because that some people were not accustomed to the restrictions which caused negative sentiment.

\subsubsection{Topic of Social-Distancing}

\begin{figure}[!htb]
  \includegraphics[width=0.99\linewidth]{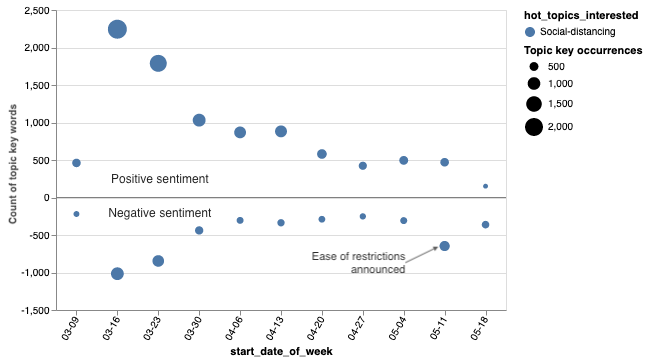}
  \caption{The dynamics of Twitter topic of ``Social-distancing''.}
\label{fig:HT_socialdist}
\end{figure}

Fig.~\ref{fig:HT_socialdist} presents the dynamics of Twitter topic of social-distancing. It shows that the topic of social-distancing appeared from the week of 9 March and significantly increased to a peak after a week. This is maybe because that the confirmed cases of COVID-19 were increased significantly and the government encouraged people to increase social-distancing, which caused a significant increase of public conversation on social-distancing on Twitter. After that week, the occurrence of the topic of social-distancing was gradually decreased until the week of 6 April and then kept stable relatively. Overall, the community showed the positive sentiment dominantly from the beginning of social-distancing until the week of 11 May. This is maybe because that the NSW government announced the ease of restrictions on 10 May as mentioned previously, which resulted in the negative sentiment to the social-distancing from the community.

\subsubsection{Topic of Jobkeeper}

\begin{figure}[!htb]
  \includegraphics[width=0.99\linewidth]{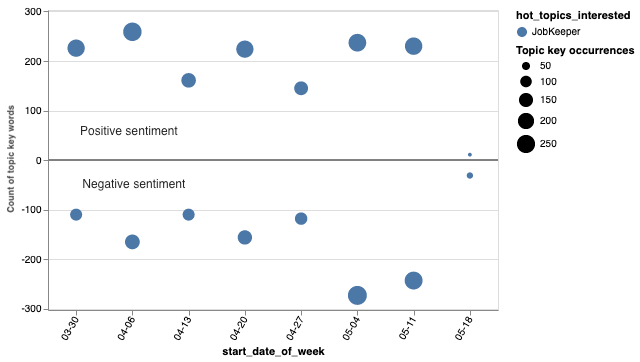}
  \caption{The dynamics of Twitter topic of ``Jobkeeper''.}
\label{fig:HT_jobkeeper}
\end{figure}

The Australian Government introduced the Jobkeeper wage subsidy program at the end of March, allowing businesses to claim a fortnightly payment of \$1,500 per eligible employee from 30 March 2020, for a maximum of 6 months\footnote{\url{https://treasury.gov.au/sites/default/files/2020-04/Fact_sheet_supporting_businesses_0.pdf}}. The starting time of jobkeeper topic was also clearly demonstrated in the dynamics of Twitter topic of jobkeeper as shown in Fig.~\ref{fig:HT_jobkeeper}. This figure shows that the public kept a positive sentiment dominantly to jobkeeper program from the beginning of its introduction until the week of 27 April. From the week of 4 May, people showed a negative sentiment dominantly. This is maybe because that people had received negative news on operations of jobkeeper program. For example, it was reported that 
the number of people on jobkeeper was revised down by 3 million due to the errors at the beginning of the program\footnote{\url{https://www.abc.net.au/news/2020-05-22/jobkeeper-numbers-cut-by-3-million-businesses-accounting-bungle/12277488}}.

\subsection{Sentiment Dynamics Affected by Events}

Community sentiment may also be affected by big events during the COVID-19 period. For example, the Ruby Princess Cruise docked in Sydney Harbour on 18 March 2020. Despite having passengers with COVID-19 symptoms, about 2700 passengers were allowed to disembark on 19 March without isolation or other measures, which was considered to create a coronavirus hotbed in Australia. The Ruby Princess Cruise has been linked to at least 662 confirmed cases and 22 deaths of COVID-19\footnote{\label{fn:au_covid_time}\url{https://www.theguardian.com/world/2020/may/02/australias-coronavirus-lockdown-the-first-50-days}}. 

\begin{figure}[!htb]
  \includegraphics[width=0.99\linewidth]{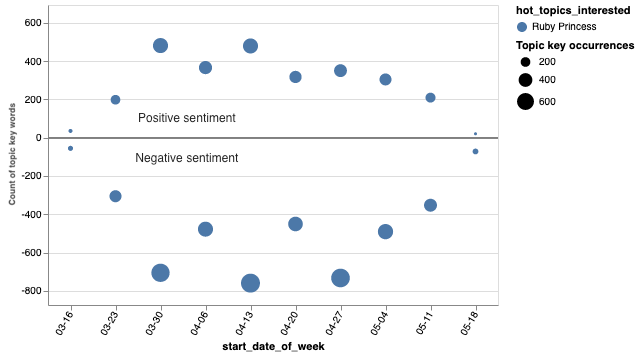}
  \caption{The dynamics of Twitter topic of ``Rubyprincess''.}
\label{fig:HT_rubyprincess}
\end{figure}

Fig.~\ref{fig:HT_rubyprincess} presents the community sentiment dynamics related to the Ruby Princess. It shows that the topic of the Ruby Princess appeared from the week of 16 March and was increased significantly after that. It reached a peak at the week of 30 March and kept high occurrences until the beginning of May. This trend is well aligned with the actual timeline of the public reporting of confirmed cases and deaths as well as other events (e.g. police in NSW announced a criminal investigation into the Ruby Princess cruise ship debacle on 5 April) related to the Ruby Princess \textsuperscript{\ref{fn:au_covid_time}}. Overall, the community showed a dominant negative sentiment to Ruby Princess throughout the event period. 

\section{Conclusions and Future Work}

This study conducted a comprehensive analysis of the sentiment dynamics in the state of New South Wales (NSW) in Australia due to the COVID-19 pandemic. 
Instead of the country-level study, the sentiment in this work was analysed at local government areas (LGAs) level based on more than 94 million tweets collected from Twitter for a 5-month period started from 1 January 2020. 
The results of the dynamical sentiment analysis showed that the overall sentimental polarity was positive in  NSW and the positive sentiment was decreased from the beginning of March to May 2020 due to the significant increase of COVID-19 confirmed cases from 8 March and the further carried out lockdown policies. 
When we drilled down into LGAs, it was found that different LGAs showed different sentiment polarity scores during the study time, and each LGA may have different sentiment polarity scores on different days. 
Despite the dominant positive sentiment most of days during the study period, some LGAs experienced significant sentiment dynamics maybe because of the serious COVID-19 infections in those LGAs. This study also analysed the sentimental dynamics delivered by the hot topics such as government policies (e.g. the Australia JobKeeper program) and the focused social events (e.g. the Ruby Princess Cruise). The results showed that the implemented policies or occurred events affected people's overall sentiment differently, and at the different stage, people showed different sentiment.

This paper presented a case study of the sentiment dynamics due to COVID-19. More interesting topics can be explored based on the current study in the future.
For example, the topic modelling techniques can be applied to conduct more comprehensive analysis to get people's topic level sentiments.
The individual level sentiment dynamics can also be analysed to help government and local communities to locate the people that may suffer from the negative sentiments. Besides the sentiment polarity, other mental health issues due to COVID-19 such as depression and anxiety will also be analysed in Twitter in our future work.

\begin{acknowledgements}
Authors would like to thank Zhangwei Chen for his help in the collection of Twitter data.
\end{acknowledgements}

\noindent \textbf{Conflict of Interest:} No conflict of interest exists for all participating authors.

%
%

\bibliographystyle{spmpsci}      
\bibliography{community_sentiment} 

\begin{thebibliography}{10}
\providecommand{\url}[1]{{#1}}
\providecommand{\urlprefix}{URL }
\expandafter\ifx\csname urlstyle\endcsname\relax
  \providecommand{\doi}[1]{DOI~\discretionary{}{}{}#1}\else
  \providecommand{\doi}{DOI~\discretionary{}{}{}\begingroup
  \urlstyle{rm}\Url}\fi

\bibitem{Anuprathibha2016}
Anuprathibha, T., Selvib, C.S.: {A survey of twitter sentiment analysis}.
\newblock IIOAB Journal \textbf{7}(9Special Issue), 374--378 (2016)

\bibitem{barkur_sentiment_2020}
Barkur, G., Vibha, Kamath, G.B.: Sentiment analysis of nationwide lockdown due
  to {COVID} 19 outbreak: Evidence from india.
\newblock Asian Journal of Psychiatry  (2020)

\bibitem{bhat_sentiment_2020}
Bhat, M., Qadri, M., Beg, N.u.A., Kundroo, M., Ahanger, N., Agarwal, B.:
  Sentiment analysis of social media response on the covid19 outbreak.
\newblock Brain, Behavior, and Immunity  (2020)

\bibitem{dubey_twitter_2020}
Dubey, A.D.: Twitter sentiment analysis during {COVID}-19 outbreak.
\newblock {SSRN} Scholarly Paper {ID} 3572023, Social Science Research Network
  (2020)

\bibitem{gibbons_twitter-based_2019}
Gibbons, J., Malouf, R., Spitzberg, B., Martinez, L., Appleyard, B., Thompson,
  C., Nara, A., Tsou, M.H.: Twitter-based measures of neighborhood sentiment as
  predictors of residential population health.
\newblock {PLoS} {ONE} \textbf{14}(7) (2019)

\bibitem{Go2009}
Go, A., Bhayani, R., Huang, L.: {Twitter Sentiment Classification using Distant
  Supervision}.
\newblock Processing \textbf{-}, 1--6 (2009)

\bibitem{Hassan2013}
Hassan, A., Abbasi, A., Zeng, D.: {Twitter sentiment analysis: A bootstrap
  ensemble framework}.
\newblock Proceedings - SocialCom/PASSAT/BigData/EconCom/BioMedCom 2013 pp.
  357--364 (2013).
\newblock \doi{10.1109/SocialCom.2013.56}

\bibitem{hutto_VADER_2014}
Hutto, C.J., Gilbert, E.: Vader: A parsimonious rule-based model for sentiment
  analysis of social media text.
\newblock In: ICWSM (2014)

\bibitem{jaidka_estimating_2020}
Jaidka, K., Giorgi, S., Schwartz, H.A., Kern, M.L., Ungar, L.H., Eichstaedt,
  J.C.: Estimating geographic subjective well-being from twitter: A comparison
  of dictionary and data-driven language methods.
\newblock Proceedings of the National Academy of Sciences \textbf{117}(19),
  10165--10171 (2020)

\bibitem{Jianqiang2018}
Jianqiang, Z., Xiaolin, G., Xuejun, Z.: {Deep Convolution Neural Networks for
  Twitter Sentiment Analysis}.
\newblock IEEE Access \textbf{6}, 23253--23260 (2018).
\newblock \doi{10.1109/ACCESS.2017.2776930}

\bibitem{Liu2012}
Liu, B.: Sentiment analysis and opinion mining.
\newblock Synthesis lectures on human language technologies \textbf{5}(1),
  1--167 (2012)

\bibitem{Mittal2012}
Mittal, A., Goel, A.: Stock prediction using twitter sentiment analysis.
\newblock Standford University, CS229 (2011 http://cs229. stanford.
  edu/proj2011/GoelMittal-StockMarketPredictionUsingTwitterSentimentAnalysis.
  pdf) \textbf{15} (2012)

\bibitem{montemurro_emotional_2020}
Montemurro, N.: The emotional impact of {COVID}-19: From medical staff to
  common people.
\newblock Brain, Behavior, and Immunity  (2020)

\bibitem{ortega2013ssa}
Ortega, R., Fonseca, A., Montoyo, A.: Ssa-uo: unsupervised twitter sentiment
  analysis.
\newblock In: Second joint conference on lexical and computational semantics (*
  SEM), vol.~2, pp. 501--507 (2013)

\bibitem{thelwall2010sentiment}
Thelwall, M., Buckley, K., Paltoglou, G., Cai, D., Kappas, A.: Sentiment
  strength detection in short informal text.
\newblock Journal of the American society for information science and
  technology \textbf{61}(12), 2544--2558 (2010)

\bibitem{noauthor_coronavirus_2020}
{World Health Organization}: Coronavirus disease (covid-19) pandemic (2020).
\newblock
  \urlprefix\url{https://www.who.int/emergencies/diseases/novel-coronavirus-2019}.
\newblock [Online; accessed 2020-05-15]

\bibitem{Yang2020}
Yang, S., Huang, G., Ofoghi, B., Yearwood, J.: {Short text similarity
  measurement using context-aware weighted biterms}.
\newblock In: Concurrency Computation. John Wiley and Sons Ltd (2020).
\newblock \doi{10.1002/cpe.5765}

\end{thebibliography}

\end{document}